\documentclass[10pt]{article}
\usepackage{bbm}
\usepackage{amsmath}
\usepackage{amsfonts,amsbsy}
\usepackage{amssymb}
\usepackage{graphicx}

\def\k{{\boldsymbol k}}

\def\x{{\boldsymbol x}}

\def\bs{\boldsymbol}

\newcommand{\tr}{\mathrm{tr}}
\newcommand{\D}{\mathcal{D}}
\newcommand{\Ham}{\mathcal{H}}
\newcommand{\Lag}{\mathcal{L}}

\newcommand{\Ac}{\mathcal{A}}
\newcommand{\Ec}{\mathcal{E}}
\newcommand{\Fc}{\mathcal{F}}

\newcommand{\ti}{t_{_I}}
\newcommand{\tf}{t_{_F}}
\newcommand{\xin}{x_{_I}}
\newcommand{\xf}{x_{_F}}
\newcommand{\xc}{x^{\text{c}}}
\newcommand{\xcf}{x^{\text{c}}_{_F}}
\newcommand{\Xin}{X_{_I}}
\newcommand{\pin}{p_{_I}}
\newcommand{\pfi}{p_{_F}}
\newcommand{\pc}{p^{\text{c}}}
\newcommand{\Pin}{P_{_I}}

\newcommand{\phiin}{\phi_{_I}}
\newcommand{\phic}{\phi^{\text{c}}}
\newcommand{\phicf}{\phi^{\text{c}}_{_F}}
\newcommand{\piin}{\pi_{_I}}
\newcommand{\pic}{\pi^{\text{c}}}

\newcommand{\rmd}{\mathrm{d}}
\newcommand{\rmi}{\mathrm{i}}
\newcommand{\rme}{\mathrm{e}}

\newcommand{\half}{\textstyle\frac{1}{2}}

\catcode`\@=11

\newcount\@tempcntc
\def\@citex[#1]#2{\if@filesw\immediate\write\@auxout{\string\citation{#2}}\fi
  \@tempcnta\z@\@tempcntb\m@ne\def\@citea{}\@cite{%
        \@for\@citeb:=#2\do%
    {\@ifundefined{b@\@citeb}%
        {\@citeo\@tempcntb\m@ne\@citea%
                \def\@citea{,\penalty\@m\ }{\bf ?}\@warning%
                {Citation `\@citeb' on page \thepage \space undefined}}%
        {\setbox\z@\hbox{\global\@tempcntc0\csname b@\@citeb\endcsname\relax}
     \ifnum\@tempcntc=\z@ \@citeo\@tempcntb\m@ne%
       \@citea\def\@citea{,\penalty\@m}%
       \hbox{\csname b@\@citeb\endcsname}%
     \else%
      \advance\@tempcntb\@ne%
      \ifnum\@tempcntb=\@tempcntc%
      \else\advance\@tempcntb\m@ne\@citeo%
      \@tempcnta\@tempcntc\@tempcntb\@tempcntc\fi\fi}}\@citeo}{#1}}%

\def\@citeo{\ifnum\@tempcnta>\@tempcntb\else\@citea
  \def\@citea{,\penalty\@m}%
  \ifnum\@tempcnta=\@tempcntb\the\@tempcnta\else
   {\advance\@tempcnta\@ne\ifnum\@tempcnta=\@tempcntb \else
\def\@citea{--}\fi
    \advance\@tempcnta\m@ne\the\@tempcnta\@citea\the\@tempcntb}\fi\fi}

\catcode`\@=12

\begin{document}

\title{\bf Initial Singularity of the Little Bang}
\author{Kenji Fukushima$^{(1)}$, Fran\c{c}ois Gelis$^{(2)}$,
 Larry McLerran$^{(1,3)}$}
\maketitle
\begin{center}
\begin{enumerate}
\item RIKEN BNL Research Center,\\
  Brookhaven National Laboratory,\\
  Upton, NY-11973, USA
\item Service de Physique Th\'eorique (URA 2306 du CNRS)\\
  CEA/DSM/Saclay, B\^at. 774\\
  91191, Gif-sur-Yvette Cedex, France
\item Department of Physics, Bldg. 510 A,\\
Brookhaven National Laboratory,\\
  Upton, NY-11973, USA
\end{enumerate}
\end{center}

\begin{abstract}
 The Color Glass Condensate (CGC) predicts the form of the nuclear
 wavefunction in QCD at very small $x$.  Using this, we compute the
 wavefunction for the collision of two nuclei, infinitesimally in the
 forward light cone.  We show that the Wigner transformation of this
 wavefunction generates rapidity dependent fluctuations around the
 boost invariant classical solution which describe the Glasma in the
 forward light cone.
\end{abstract}

\begin{flushright}
{\small\sf RBRC-621\\ BNL-NT-06/41\\ SPhT-T06/139}
\end{flushright}


\section{Introduction}

  The Color Glass Condensate (CGC) provides a description of the
wavefunction of a hadron at very small values of
$x$~\cite{mv,jkmw,jklw,im2001,ilm}.  The CGC is a high density state
of gluons which is controlled by a weak coupling, due to the high
gluon density.  Its properties are computable from first principles in
QCD, at least in the limit of extremely high density.

  The CGC has also been applied to heavy ion collisions in order to
generate the initial conditions for the evolution of matter in the
forward light cone~\cite{mkw,kv,knv,lappi1,lappi}.  There are boost
invariant solutions of the equations of motion supposedly appropriate
for the high energy limit.  The matter in the forward light cone,
which we call the Glasma~\cite{lappi}, initially has large
longitudinal color electric and magnetic fields, and as well as a
large value of the Chern-Simons charge density.  As time evolves,
transverse color electric and magnetic fields appear as the remnants
of the decaying longitudinal fields.  This occurs as a consequence of
the classical equations of motion.  Eventually as the system becomes
very dilute, these transverse color electric and magnetic fields may
be treated as gluons, which form a quark-gluon plasma.  The solution
to the classical equations of motion is boost invariant and describes
a system of expanding classical fields.

  It has recently been discovered that the boost invariant solution of
the equations of motion are unstable with respect to rapidity
dependent perturbations~\cite{roma}. This instability has in fact
close connections with the Weibel instabilities encountered in the
physics of anisotropic plasmas~\cite{mro,arnold,roma1}, and it is
speculated that such an instability may help the system created after
heavy ion collisions reach a state of local equilibrium
\cite{moore,arnold1,bodeker,dumitru,dumitru1,strickland,mro1}.  Whether
or not such instabilities can grow to sufficient magnitude as to
become as large as the classical solution is the subject of current
investigations.  One piece of the puzzle which is not yet understood
is the spectrum of the initial fluctuations.  It is the purpose of
this paper to derive an expression for the probability distribution of
these fluctuations.

  If such fluctuations can grow to a magnitude comparable to the
classical boost invariant field, then one has amplified initial
quantum fluctuations to macroscopic chaotic turbulence.  Such
turbulence may ultimately be responsible for producing a
\textit{thermalized} Quark Gluon Plasma.  Indeed, it proves difficult to
thermalize an expanding system of quarks and gluons by conventional
multiple scattering formulae~\cite{son}.

\begin{figure}[ht]
 \begin{center}
   \includegraphics[width=1.0\textwidth]{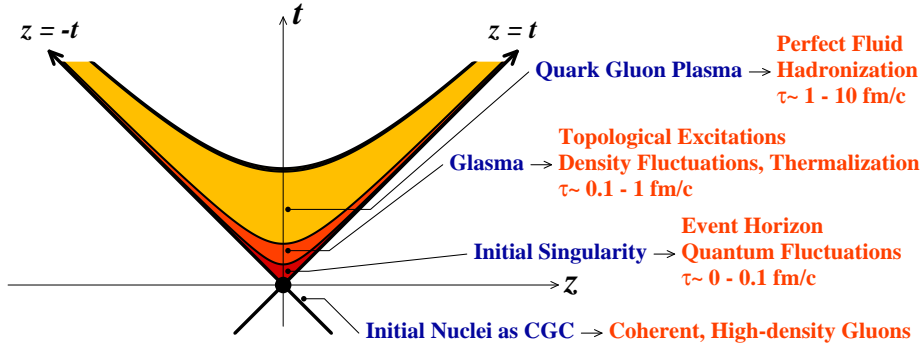}
      \caption{A light cone diagram illustrating the evolution of
               matter produced in heavy ion collisions }
      \label{lbang}
 \end{center}
\end{figure}

  The evolution from the initial collision to the final state is shown
in Fig.~\ref{lbang}.  Because of the similarities between the
expansion of the universe in cosmology and the expansion of matter in
heavy ion collisions, we call the latter the ``little bang''.  The
initial singularity of cosmology is replaced by the singularity in the
classical equations of motion associated with the collision.  As in
cosmology, where topological transitions associated with Chern-Simons
charge may be responsible for generating baryon number, during the
Glasma phase, there are also topological helicity flip
transitions~\cite{lappi,alex}.  It is during this time that
instabilities develop, and there may be a Kolmogorov spectrum of
density fluctuations generated~\cite{arnold1,dumitru1,mueller}.  This
spectrum is similar to the spectrum of density fluctuations generated
during inflationary cosmology.  After inflation, the system reheats
and thermalizes forming an electroweak plasma.  In the little bang,
thermalization might also be achieved after the Glasma expansion.  Of
course such an analogy between the little bang is not perfect, and it
has not been established that the Glasma can in fact become
thermalized.  Nevertheless, in the little bang, all of the physics is
in principle understood from QCD in a small coupling regime, and one
should be able ultimately to compute the evolution of the system.

  We begin this paper by a discussion of the evolution of an unstable
system in quantum mechanics.  We discuss a Gaussian initial
wavefunction in the path-integral representation in which a
description equivalent to the Schwinger-Keldysh formalism~\cite{ms}
naturally arises.  We show that the distribution of initial
fluctuations is given by the Wigner transform of the quantum
mechanical wavefunction~\cite{jeon}.  These initial fluctuations are
further evolved by solving classical equations of motion.  We then
proceed to show that the same procedure works in quantum field theory.
We then compute the quantum field theoretical wavefunction which
describes the collision of two hadrons, at a time infinitesimally in
the forward light-cone, after the collision.  This should provide a
practical algorithm for the computation of the evolution of matter
produced in very high energy heavy ion collisions.

  The necessity of summing over the initial quantum fluctuations when
solving the classical Yang-Mills equations also appears when one goes
beyond the classical approximation for the description of the fields
in the forward light-cone. The formalism necessary for calculating
loop corrections in a regime of strong sources and fields has been
developed in Ref.~\cite{gv} (and applied to the computation of the
production of quark pairs -- the simplest among the NLO contributions
-- in Ref.~\cite{gkl}). In this systematic approach, the instability
of the boost invariant classical solution manifests itself as a
divergence in the one-loop correction to the inclusive gluon spectrum,
and it seems that the resummation of these divergent terms leads
directly to an average over fluctuations of the classical initial
conditions~\cite{glv}.

  The description we advocate here has also some philosophical
similarities with the work of Kharzeev {\it et al.}, who make an
analogy to Hawking radiation~\cite{klt}.  We are unable to identify a
Hawking temperature in our computation, but the idea that there is an
initial spectrum of fluctuations generated at the initial singularity
due to a quantum treatment of this singularity is similar.


\section{The Case of Quantum Mechanics}


\subsection{General Formulation}

  We shall consider the time evolution of a generic quantum mechanical
system from the initial time $\ti$ to the final time $\tf$.  If the
system is as simple as an harmonic oscillator, the quantum
fluctuations at the final time $\tf$ can be expressed in terms of the
initial quantum fluctuation at $\ti$ which evolve according to the
classical equations of motion.  In more complicated situations beyond
the harmonic oscillator problem, the formulation we develop in what
follows still holds unchanged, as long as we can make a Gaussian
approximation in the weak coupling regime.

\begin{figure}[ht]
\begin{center}
 \includegraphics[width=0.60\textwidth]{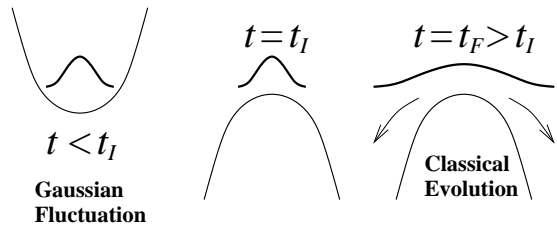}
 \caption{Schematic picture of the time evolution of quantum
 fluctuations.  The ground state of a steady system for $t<\ti$ has
 a Gaussian distribution of quantum fluctuations.  The distribution
 at a later time $\tf$ is given by the classical evolution from the
 initial distribution at $t=\ti$.}
 \label{fig:sketch}
\end{center}
\end{figure}

  Let us consider a simple example~\cite{barton}.  Supposing that we
have a convex parabolic potential as sketched in Fig.~\ref{fig:sketch}
until the initial time $\ti$, we know that the ground state has a
Gaussian distribution of quantum fluctuations around the minimum of
the potential.  Let us now assume that the potential suddenly becomes
inverted at $\ti$~: then the system is subject to an instability.  If
the problem was purely classical, the unstable state would not decay
at all provided it lies exactly on the top of the potential curve.
The decay into stable states is however unavoidable in quantum
mechanics because of quantum fluctuations around the minimum.  As long
as higher order quantum fluctuations are negligible, it should be
acceptable to approximate the time evolution classically under such a
potential causing the instability.  As a result, we can anticipate
that the convolution of the initial dispersion and the classical
evolution provides the later distribution of quantum fluctuations at
$\tf$, and we shall verify this in the forthcoming discussion.  We
would like to emphasize here that our formulation is not limited to
this sort of specific instability problem, but applicable to more
generic problems.

  In the language of quantum physics the ground state wavefunction at
$\ti$ embodies the zero-point oscillation.  The position and momentum
fluctuate according to the uncertainty principle.  The expectation
value of physical observables is defined as an average weighted by the
wavefunction.  Let us consider an observable $\mathcal{O}(\xf)$ which
is a function of the position $\xf$ at the final time $\tf$.  [We
specifically denote the position variable at $\ti$ and $\tf$ as $\xin$
and $\xf$ respectively in order to distinguish them from the position
at other times.]  By definition the expectation value at $\tf$ is
given by
\begin{equation}
 \langle\mathcal{O}\rangle_{\tf} = \int\rmd\xf\, \left|\psi(\xf,\tf)
  \right|^2\,\mathcal{O}(\xf)\;.
\label{eq:ev_qm}
\end{equation}
One can easily generalize the above expression to a situation
involving observables that depend not only on the position $\xf$ but
also on the momentum $\pfi$.  Using $\pfi=-\rmi\partial/\partial\xf$
and $\mathcal{O}(\xf,-\rmi\partial/\partial\xf)$ acting on
$\psi(\xf,\tf)$, one can reach the same conclusion as in
Eq.~(\ref{eq:formula_qm}) in what follows, but for simplicity we only
perform the derivation for the case of Eq.~(\ref{eq:ev_qm}).

  The path-integral formulation naturally provides us the relation
between the initial and final wavefunctions,
\begin{equation}
 \psi(\xf,\tf)=\int\rmd\xin\,\big[\D x(t)\,\D p(t)\big]\, 
\psi(\xin,\ti)\,\exp\biggl\{
  \,\rmi\int_{\ti}^{\tf}\!\rmd t\,\bigl[p\dot{x}-H(p,x)\bigr] \biggr\}\;,
\label{eq:path_qm}
\end{equation}
where the integrals over the paths $x(t)$ and $p(t)$ extend from $\ti$
to $\tf$, with the boundary condition $x(\ti)=\xin$ and $x(\tf)=\xf$.
The Hamiltonian is denoted by $H(p,x)$.  We will perform the
path-integral approximately in order to find an analytically manageable
expression.

  We expand the phase in the exponential around the stationary point,
in order to reduce Eq.~(\ref{eq:path_qm}) to Gaussian functional
integrals.  The stationary condition leads to the classical path
$x=\xc(t)$ and $p=\pc(t)$ determined by Hamilton's equations of
motion,
\begin{equation}
 \dot{\xc}=\frac{\partial H(\pc,\xc)}{\partial\pc}, \qquad
 \dot{\pc}=-\frac{\partial H(\pc,\xc)}{\partial\xc}
\label{eq:eom_qm}
\end{equation}
with the initial conditions $\xc(\ti)=\xin$ and $\pc(\ti)=\pin$.  We
note that the final position $\xcf(\xin,\pin)\equiv\xc(\tf)$ is unique
with $\xin$ and $\pin$ given, which means that we must consider the
initial momentum $\pin$ as a function of $\xin$ and $\xf$.  Naturally,
the quantum fluctuation around the classical path must be restricted
at $\ti$ and $\tf$ to satisfy the boundary condition $\delta
x(\ti)=\delta x(\tf)=0$. Up to the quadratic order in quantum
fluctuations, therefore, we have
\begin{equation}
 \begin{split}
 & \psi(\xf,\tf)= \int\rmd\xin\,\psi(\xin,\ti)\,
  \exp\biggl\{\,\rmi\int_{\ti}^{\tf}\!\rmd t\bigl[\pc\dot{\xc}
  -H(\pc,\xc)\bigr]\biggr\} \\
 & \qquad\qquad \times\!\! \!\!\int
  \limits_{\delta x(\ti)=\delta x(\tf)=0} \!\!\!\!\!\!\!\!\!\!\!\!
  \big[\D\delta x(t)\,\D\delta p(t)\big] \,
  \exp\biggl\{\,\rmi\int_{\ti}^{\tf}\!\rmd t\Bigl[
  \delta p\,\delta\dot{x} \\
 &\qquad\qquad\qquad\qquad\qquad
  -\frac{\partial^2 H}{\partial p^2}\delta p^2
  \!-\!\frac{\partial^2 H}{\partial x^2}\delta x^2
  \!-\!2\frac{\partial^2 H}{\partial p\,\partial x}\delta p\,\delta x
  \Bigr]\biggr\}\;.
 \end{split}
\end{equation}
In the expansion of the argument of the exponential, the linear order
terms in the fluctuations vanish because of the classical equations of
motion.  The first integral in the above expression is the classical
part corresponding to the WKB approximation, where the $\xf$
dependence is implicit through the fact that $\xc(t)$ is the classical
trajectory that starts at $\xin$ and ends at $\xf$.  The second
integral represents the one-loop quantum corrections leading to the
determinant of the propagator with the Dirichlet boundary condition.
It is obvious from the above expression that, if $H$ is the
Hamiltonian of an harmonic oscillator (therefore consisting only of
quadratic terms in $x$ and $p$), e.g., $H=\half p^2+\half\omega^2
x^2$, the WKB approximation is exact.  It is because none of
$\partial^2 H/\partial p^2\!=\!1$,
$\partial^2 H/\partial x^2\!=\!\omega^2$, and
$\partial^2 H/\partial p\,\partial x\!=\!0$ depends on $\xin$ nor
$\pin$, and thus the integration with respect to fluctuations merely
produces an irrelevant purely numerical factor.  

  In general, the one-loop integral gives $[\det G^{-1}(\xc)]^{-1/2}$
where $G$ is the propagator of a fluctuation in the presence of the
classical background $\xc$.  Usually, this determinant has no explicit
$\pc$ dependence since usual quantum theories have at most quadratic
terms with respect to the canonical momenta.  These quantum
corrections alter $H(\pc,\xc)$ to be $H(\pc,\xc)-\frac{\rmi}{2}\ln
G^{-1}(\xc)$ and can be seen as an effective potential in the
classical Hamiltonian.

{}From now on we assume that the $\xc$ dependence in $H(\pc,\xc)$ is
so strong that we can drop the $\xc$ dependence from the quantum
corrections $-\frac{\rmi}{2}\ln G^{-1}(\xc)$.  This assumption should
be checked case by case.  Generally speaking, this is acceptable in
instability problems like the one illustrated in
Fig.~\ref{fig:sketch}.  The quantum corrections modify the tree-level
potential, which would eventually affect the equations of motion.  But
if the tree-level potential is steep enough, the alteration due to
quantum corrections should be negligible.  Hence, our treatment is
equivalent to assuming that the time evolution is dominantly governed
by the tree-level potential.

  The probability distribution for the positions at final time is
therefore calculated with the classical parts alone as
\begin{equation}
 \begin{split}
 &\left|\psi(\xf,\tf)\right|^2
  =\int\rmd\xin'\,\rmd\xin\,\psi^\ast(\xin',\ti')\psi(\xin,\ti) \\
 & \qquad\qquad
\times \exp\biggl\{\,\rmi
  \int_{\ti}^{\tf}\rmd t\,\Bigl[\bigl(\pc\dot{\xc}\!-\!H(\pc,\xc)
  \bigr)\!-\!\bigl({\pc}'\dot{\xc}'\!-\!H({\pc}',{\xc}')\bigr)
  \Bigr]\biggr\}\;.
 \end{split}
\end{equation}
The phase difference induced by the different initial conditions,
$\xin$ and $\xin'$, can be simplified thanks to Hamilton's equations
of motion (\ref{eq:eom_qm}) as
\begin{align}
 &\int_{\ti}^{\tf}\rmd t\,\Bigl[\bigl(\pc\dot{\xc}\!-\!H(\pc,\xc)
  \bigr)\!-\!\bigl({\pc}'\dot{\xc}'\!-\!H({\pc}',{\xc}')\bigr)
  \Bigr] \notag\\
 &\qquad
=\int_{\ti}^{\tf}\rmd t
  \int_{\{{\pc}'(t),{\xc}'(t)\}}^{\{\pc(t),\xc(t)\}}\biggl\{
  \rmd\pc\dot{x}+\pc\rmd\dot{\xc}
  -\frac{\partial H}{\partial\pc}\rmd\pc
  -\frac{\partial H}{\partial\xc}\rmd\xc\biggr\} \notag\\
 &\qquad
=-\int_{\xin'}^{\xin}\rmd\tilde{\xin}\;\pin(\tilde{\xin},\xf)\;.
\label{eq:phase-diff}
\end{align}
{}From the middle to the last line only the surface term of the time
integration remains.  There is no finite contribution from the $t=\tf$
side because $\xf$ is common.  It should be noted that
$\pin(\xin,\xf)$ in the last line is a \textit{function} which gives
the value of the initial momentum $\pin$ as a function of the initial
and final positions, i.e. $\pin(\xin,\xf)=\pin$ and
$\pin(\xin',\xf)=\pin'$.  Here, in order to get a convenient
expression, we shall define $\Pin$ as the initial momentum
corresponding to the average of the two initial positions (and the
same final position $\xf$) by
\begin{equation}
 \Pin \equiv \pin(\Xin,\xf) \;,
\end{equation}
where $\Xin\equiv\half(\xin+\xin')$.  Then we can write the last line
of Eq.~(\ref{eq:phase-diff}) as follows,
\begin{align}
 -\int_{\xin'}^{\xin}\rmd\tilde{\xin}\;\pin(\tilde{\xin},\xf)
 = & -\int_{-\half\delta\xin}^{\half\delta\xin}\rmd\Delta\tilde{\xin}\;
  \pin(\Xin+\Delta\tilde{\xin},\xf) \notag\\
 \simeq & -\Pin\delta\xin-\frac{1}{24}\frac{\partial^2\pin(\Xin,\xf)}
  {\partial\Xin^2}\delta\xin^3+\cdots
\end{align}
where we defined $\delta\xin\equiv\xin-\xin'$.  Within the Gaussian
approximation that we will employ later, it is enough for us to keep
only the first term in the above expansion.

  The expectation value at the final time is now transformed in the
WKB approximation from Eq.~(\ref{eq:ev_qm}) into
\begin{equation}
 \langle\mathcal{O}\rangle_{\tf} =\int\rmd\xf\,\rmd\xin'\,\rmd\xin\;
  \psi^\ast(\xin',\ti)
  \psi(\xin,\ti)\;\rme^{-\rmi \Pin\delta\xin}\,\mathcal{O}(\xf) \;.
\end{equation}
Here we can replace the $\xf$ integration by an integration over
$\Pin$. Indeed, $\xf$, $\Pin$ and $\Xin$ are related by the fact that
the classical path that starts at the initial position $\Xin$ and with
the initial momentum $\Pin$ ends at the position $\xf$,
i.e. $\xf=\xcf(\Xin,\Pin)$.  Such a replacement is accompanied by the
Jacobian $|\partial\xcf/\partial\Pin|$ as a function of $\Xin$,
$\Pin$, and $\tf$.  One can, in principle, obtain the Jacobian once
one solves the classical path $\xc(\xin,\pin)$.  This is, however,
only a prefactor of the exponential for the phase of which we made the
stationary-point approximation.  It should be, therefore, consistent
to drop the Jacobian prefactor off within the WKB approximation.

  After all, rewriting $\xin=\Xin+\half\delta\xin$ and
$\xin'=\Xin-\half\delta\xin$ and integrating with respect to
$\delta\xin$, we formally reach the following expression,
\begin{align}
 &\langle\mathcal{O}\rangle_{\tf} \notag\\
 =&\int\rmd\Xin\,\rmd\delta\xin\,\rmd\Pin\,\psi^\ast(\Xin
  -\half\delta\xin,\ti)\psi(\Xin+\half\delta\xin,\ti)\,
  \rme^{-\rmi\Pin\delta\xin}\,\mathcal{O}\bigl(\xcf(\Xin,\Pin)\bigr)
  \notag\\
 =&\int\rmd\Xin\,\rmd\Pin\;W(\Xin,\Pin)\;\mathcal{O}
  \bigl(\xcf(\Xin,\Pin)\bigr)\;,
\label{eq:formula_qm}
\end{align}
where $W(\Xin,\Pin)$ is the Wigner transform of the product of the two
initial wavefunctions, defined as
\begin{equation}
 W(\Xin,\Pin)\equiv\int\rmd\delta\xin\,\psi^\ast(\Xin-\half\delta\xin,
  \ti)\psi(\Xin+\half\delta\xin,\ti)\,\rme^{-\rmi\Pin\delta\xin}\;.
\end{equation}
This is our formula and the purpose of our work is to calculate the
Wigner function analytically for the problem of heavy ion collisions,
which should be a necessary input for numerical studies provided that
the classical equations of motion are solved.


\subsection{Simple Example}

  It would be instructive to demonstrate how the formula
(\ref{eq:formula_qm}) works for an harmonic oscillator for which
explicit calculations are feasible.  Let us estimate the average of
the observable ${\mathcal{O}}\equiv \xf^2$ in a direct computation and
by means of Eq.~(\ref{eq:formula_qm}) and then compare the results.
As we have already mentioned, the WKB approximation is exact in the
case of the harmonic oscillator problem, which means that the two
results should coincide.  This is a trivial check of the validity of
our formula (\ref{eq:formula_qm}).

  We shall characterize the initial state as a Gaussian~\cite{barton};
\begin{equation}
 \psi(\xin,\ti=0) = \bigl(\pi^{\frac{1}{2}}b\bigr)^{-\frac{1}{2}}
  \exp\bigl(-\xin^2/2b^2\bigr)\; .
\label{eq:example_gauss}
\end{equation}
The Green's function in the inverted harmonic oscillator with the
Hamiltonian $H=\half p^2-\half\omega^2 x^2$, which is defined by
\begin{align}
 & \bigl(\rmi\partial_t+\partial_x^2+\omega^2 x^2\bigr)\,
  G(x,x^\prime;t)=\delta(t)\delta(x-x^\prime)\; ,\notag\\
 & \lim_{t\to 0^+}G(x,x^\prime;t) = \delta(x-x^\prime)\; ,
\end{align}
is known~\cite{barton} to be 
\begin{equation}
 G(x,x';t) = \bigl(2\pi\rmi\sinh\omega t\bigr)^{-\frac{1}{2}}
  \exp\biggl\{\frac{\rmi}{2\sinh\omega t}\Bigl[(x^2+x^{\prime 2})
  \cosh\omega t - 2xx'\Bigr]\biggr\} \;,
\end{equation}
from which we can directly calculate the wavefunctions at later time
by
\begin{equation}
\psi(\xf,\tf)=\int\rmd\xin\;G(\xf,\xin;\tf)\psi(\xin,0)\; .
\end{equation}
  What we are calculating is the expectation value of the position
dispersion when the potential is inverted at the initial time $\ti=0$.
After some algebraic procedures we find,
\begin{equation}
 \bigl|\psi(\xf,\tf)|^2 = \bigl[\pi^{\frac{1}{2}}B^2(\tf)\bigr]^{-1}
  \exp\bigl[-\xf^2/B^2(\tf)\bigr],
\end{equation}
where $B^2(\tf)\equiv b^2\cosh^2\omega\tf + b^{-2}\sinh^2\omega\tf$.
Then, it is straightforward to compute the average value of
${\mathcal{O}}=\xf^2$,
\begin{equation}
 \langle{\mathcal{O}}\rangle_{\tf} = \int\rmd\xf\,\bigl|\psi(\xf,\tf)\bigr|^2
  \xf^2 = \frac{1}{2}B^2(\tf) \;.
\end{equation}
This result indicates that the dispersion of the position grows
exponentially with increasing time under the inverted potential just
as sketched in Fig.~\ref{fig:sketch}.

  Let us next consider the same problem according to the formula
(\ref{eq:formula_qm}).  The Wigner function obtained from
Eq.~(\ref{eq:example_gauss}) is
\begin{equation}
 W(\Xin,\Pin) = \frac{1}{\pi}
  \exp\bigl(-\Xin^2/b^2-b^2\Pin^2\bigr) \;,
\label{eq:W_qm}
\end{equation}
where we have adjusted the normalization factor, though it is only an
unimportant prefactor, so that
$\int\rmd\Xin\,\rmd\Pin\,W(\Xin,\Pin)=1$.  The classical path is fixed
by Hamilton's equations of motion with the initial inputs
$\xc(0)=\Xin$ and $\pc(0)=\Pin$, leading to the final point at $t=\tf$
as a function of $\Xin$ and $\Pin$,
\begin{equation}
 \xcf(\Xin,\Pin) = \Xin\cosh\omega\tf + \frac{\Pin}{\omega}
  \sinh\omega\tf \;.
\end{equation}
After performing Gaussian integrations we can readily arrive at the
expectation value,
\begin{equation}
 \langle\mathcal{O}\rangle_{\tf} = \int\rmd\Xin\,\rmd\Pin\,
  W(\Xin,\Pin)\,{\xcf}^2(\Xin,\Pin) = \frac{1}{2}B^2(\tf) \;.
\end{equation}
We see that our formula perfectly reproduces the result from the
direct evaluation in this simple case of the harmonic oscillator.  The
merit of our method is that we do not have to calculate the Green's
function and the wavefunction at the final time directly.


\section{Generalization to Quantum Field Theories}
  It is not difficult to extend the formula (\ref{eq:formula_qm})
described in Quantum Mechanics to the case of Quantum Field Theories.
Let us consider in this section the case of a real scalar field theory.


\subsection{Introduction}

  For a generic scalar field theory defined by a Hamiltonian density
$\Ham[\phi,\pi]$ with field $\phi$ and canonical momentum $\pi$, we
expect that the counterpart of Eq.~(\ref{eq:formula_qm}) is
\begin{equation}
 \langle\mathcal{O}\rangle_{\tf}= \int\big[\rmd\phiin(\x)\big]\,
  \big[\rmd\piin(\x)\big]\;
 W[\phiin,\piin]\;\mathcal{O}\big[\phicf[\phiin,\piin]\big]\; ,
\end{equation}
where $\phicf[\phiin,\piin]$ is the solution of the classical
equations of motion at time $\tf$ with the initial conditions
$\phic(\ti,\x)=\phiin(\x)$ and $\pic(\ti,\x)=\piin(\x)$.  This
classical field is a solution of Hamilton's equations of motion,
\begin{equation}
 \partial_t \phic=\frac{\delta H[\pic,\phic]}{\delta\pic}\; ,
  \qquad
 \partial_t \pic=-\frac{\delta H[\pic,\phic]}{\delta\phic}\; .
\end{equation}
Here $H$ denotes the integrated Hamiltonian
$H\equiv\int\rmd^3\x\,\Ham$.  The Wigner function is defined in
the same way as in Quantum Mechanics by
\begin{equation}
 W[\phiin,\piin] \equiv \int\big[\rmd\delta\phiin(\x)\big]\;
  \Psi^\ast[\phiin-\half\delta\phiin,\ti]\,
  \Psi[\phiin+\half\delta\phiin,\ti]\,
  \rme^{-\rmi\int\!\rmd^3\x\, \piin(\x)\delta\phiin(\x)}
\end{equation}
in terms of the initial wavefunction $\Psi[\phiin,\ti]$.  Similar
expressions could be written for gauge fields after unphysical
redundant degrees of freedom are removed by fixing the gauge. In the
next subsection we will explain how one can construct the initial
wavefunction $\Psi[\phiin,\ti]$ in practice.


\subsection{Wavefunction from the Annihilation Operator}

  Let us consider the wavefunction and the associated Wigner function
for the free real scalar field theory, which is the simplest extension
to a Quantum Field Theory of the harmonic oscillator that we
considered previously in Quantum Mechanics.  In the Hamiltonian
formulation, one can express the field operator $\hat{\phi}(\x)$ and
its canonical momentum operator $\hat{\pi}(\x)$ in terms of the
creation and annihilation operators as
\begin{align}
 \hat{\phi}(\x) &=\int\frac{\rmd^3\k}{(2\pi)^3}
  \frac{1}{\sqrt{2\omega_k}}\bigl(\hat{a}_\k\,
  \rme^{\rmi\k\cdot\x} + \hat{a}_\k^\dagger\,
  \rme^{-\rmi\k \cdot\x}\bigr)\; ,\\
 \hat{\pi}(\x) &= -\rmi\int\frac{\rmd^3\k}{(2\pi)^3}
  \sqrt{\frac{\omega_k}{2}}\bigl(\hat{a}_\k\,
  \rme^{\rmi\k\cdot\x} - \hat{a}_\k^\dagger\,
  \rme^{-\rmi\k\cdot\x}\bigr)\; ,
\end{align}
in the Schroedinger representation in which operators are time
independent.  The dispersion relation for the free theory is
$\omega_k=\sqrt{\k^2+m^2}$.  The canonical quantization condition is
$[\hat{\phi}(\x),\hat{\pi}(\x')] =\rmi\delta^{(3)}(\x-\x')$ or
equivalently $[\hat{a}_\k,\hat{a}_{\k'}^\dagger]
=(2\pi)^3\delta^{(3)}(\k-\k')$.  In momentum space, we have
\begin{align}
 \hat{\phi}(\k) &\equiv\int\rmd^3\x\;\hat{\phi}(\x)\;
  \rme^{-\rmi\k\cdot\x}=\frac{1}{\sqrt{2\omega_k}}
  \bigl(\hat{a}_\k+\hat{a}_{-\k}^\dagger\bigr)\; , \\
 \hat{\pi}(\k) &\equiv\int\rmd^3\x\;\hat{\pi}(\x)\;
  \rme^{-\rmi\k\cdot\x}=-\rmi\sqrt{\frac{\omega_k}{2}}
  \bigl(\hat{a}_\k-\hat{a}_{-\k}^\dagger\bigr)\; ,
\end{align}
for which the canonical commutation relation is deduced as
$[\hat{\phi}(\k),\hat{\pi}(\k')]
=\rmi(2\pi)^3\delta^{(3)}(\k-\k')$. Let us consider an initial state
which is an eigenstate of the operator $\hat{\phi}(\k)$,
\begin{equation}
\hat{\phi}(\k)\,\Psi[\phiin,\ti]=\phiin(\k)\;\Psi[\phiin,\ti]\; .
\end{equation}
{}From the commutator between $\hat{\phi}(\k)$ and $\hat{\pi}(\k')$,
we see that $\hat{\pi}(\k)$ can be represented as the differential
operator with respect to the eigenvalue $\phiin(-\k)$, that is,
\begin{equation}
 \hat{\pi}(\k)\,\Psi[\phiin,\ti]=-\rmi\frac{\delta}
  {\delta\phiin(-\k)}\,\Psi[\phiin,\ti]\; .
\end{equation}
Here the functional derivative in momentum space is understood in our
convention as
$\delta f(\k)/\delta f(\k')=(2\pi)^3\delta^{(3)}(\k-\k')$.
The ground state is the vacuum defined by the annihilation operator,
i.e.
\begin{equation}
\hat{a}_\k\Psi[\phiin,\ti]=0\; .
\end{equation}
Thus, the ground state should follow from
\begin{equation}
 \hat{a}_\k \Psi[\phiin,\ti] = \frac{1}{2}\biggl[\sqrt{2\omega_\k}\,
  \phiin(\k)+\sqrt{\frac{2}{\omega_k}}\,
  \frac{\delta}{\delta\phiin(-\k)}\biggr]\Psi[\phiin,\ti]=0\;.
\end{equation}
It is easy to find the solution to the above equation being
\begin{equation}
 \Psi[\phiin,\ti] \propto \biggl(\prod_\k \sqrt{\frac{\omega_k}{2\pi}}
  \biggr)\exp\biggl[-\int\frac{\rmd^3\k}{(2\pi)^3}\,
  \frac{\omega_k}{2}\,\phiin(-\k)\phiin(\k)\biggr].
\label{eq:wf_sc}
\end{equation}
The prefactor is the normalization for the Gaussian function, which is
in fact unimportant in our approximation.  The meaning of this
wavefunction is that the ground state contains fluctuations with an
energy $\omega_k/2$, that is, the zero-point energy. Then the Wigner
function is
\begin{equation}
 W[\phiin,\piin]=N\exp\biggl[-\int\frac{\rmd^3\k}{(2\pi)^3}\,
  \biggl\{\omega_k\,\phiin(-\k)\phiin(\k)
  +\frac{1}{\omega_k}\,\piin(-\k)\piin(\k)\biggr\}
  \biggr]
\end{equation}
with a normalization prefactor $N$ independent of the fields and
energy.  One can clearly see that the result in the free real scalar
field theory is just the product of the Wigner functions for an
infinite assembly of independent harmonic oscillators (see
Eq.~(\ref{eq:W_qm})).


\subsection{Wavefunction from the Schr\"{o}dinger Equation}

  The strategy we used in the previous subsection, based on the
canonical approach, is obvious but not very suitable for more
complicated problems like those encountered in gauge theories.  We
will develop here an alternate method in order to find the ground
state wavefunction.

  In Quantum Mechanics, the wavefunction can usually be obtained as a
solution of the Schr\"{o}dinger equation when the Hamiltonian is
provided.  We will follow this procedure in the free real scalar field
theory and will find a wavefunction that is exactly the same as what
we have obtained in Eq.~(\ref{eq:wf_sc}). Let us begin with the
Lagrangian density defining the free real scalar field theory,
\begin{equation}
 \Lag = \frac{1}{2}(\partial_t\phi)(\partial_t\phi)
  -\frac{1}{2}(\partial_i\phi)(\partial_i\phi) -\frac{1}{2}m^2\phi^2\;.
\end{equation}
The canonical momentum is
$\pi=\delta L/\delta(\partial_t\phi)=\partial_t\phi$ (where $L$ is the
integrated Lagrangian $L\equiv\int\!\rmd^3\x\,\Lag$) and the
Hamiltonian density is by definition,
\begin{equation}
 \Ham \equiv \pi\dot{\phi}-\Lag = \frac{1}{2}\pi\pi + \frac{1}{2}
  (\partial_i\phi)(\partial_i\phi) + \frac{1}{2}m^2\phi^2 \;.
\end{equation}
Of course, we could have started directly with this Hamiltonian
density, but using the Lagrangian density as our starting point is
more appropriate when we work with more general system of coordinates.
This is so because the canonical momentum varies according to the
choice of the time variable and so does the Hamiltonian density.

{}From the equal-time commutation relation,
$[\hat\phi(t,\x),\hat\pi(t,\x')]=\rmi\delta^{(3)}(\x-\x')$, we see
that the action of the momentum operator $\hat\pi(\x)$ on an
eigenstate of $\hat\phi(\x)$ is identical to that of the derivative
$-\rmi\delta/\delta\phi(\x)$.  The Schr\"{o}dinger equation,
$\rmi\partial_t\Psi=H\Psi$, can be expressed as a functional
differential equation by means of this identification.  The time
dependence is actually separable by an ansatz $\Psi=\rme^{-\rmi
Et}\Psi_0$ where $\Psi_0$ is $t$ independent.  Then, the time
independent Schr\"{o}dinger equation arises as
\begin{equation}
 \int\rmd^3\x\,\biggl[-\frac{1}{2}\frac{\delta^2}{\delta\phi(\x)^2}
  +\frac{1}{2}\phi(\x)\bigl(-\partial_i\partial_i+m^2\bigr)\phi(\x)
  \biggr]\Psi_0 = E\Psi_0 \;.
\end{equation}
The ground state solution to this equation is a Gaussian,
\begin{equation}
 \begin{split}
 \Psi_0[\phiin] = & N\exp\biggl[-\frac{1}{2}\int\rmd^3\x\,\phiin(\x)
  \sqrt{-\partial_i\partial_i+m^2}\phiin(\x)\biggr] \\
 = & N\exp\biggl[-\int\frac{\rmd^3\k}{(2\pi)^3}\frac{\omega_k}{2}
  \phiin(-\k)\phiin(\k)\biggr] \;,
 \end{split}
\end{equation}
which is identical to the wavefunction (\ref{eq:wf_sc}). Note that the
energy eigenvalue $E$ is given by
\begin{equation}
 E = \int\rmd^3\x\int\frac{\rmd^3\k}{(2\pi)^3}
  \frac{\sqrt{k^2+m^2}}{2}\;,
\end{equation}
which is nothing but the diverging zero-point energy.  In principle,
one could also write down the wavefunctions for excited states in
terms of the Hermite polynomials.


\subsection{Wavefunction in terms of the $\tau,\eta$ coordinates}

  For later convenience we shall see the same problem in the
$\tau,\eta$ coordinates defined by
\begin{equation}
 x^0 + x^3 = \tau \rme^{\eta},\qquad
 x^0 - x^3 = \tau \rme^{-\eta}\; .
\label{eq:tau}
\end{equation}
These coordinates are a natural choice for the purpose of describing
the space-time geometry of a collisions between two high energy
projectiles.  Commonly $\tau$ and $\eta$ are called the proper time
and the rapidity respectively.  The Lagrangian density in this
coordinate system is
\begin{equation}
 \tau\Lag = \frac{\tau}{2} (\partial_\tau\phi)(\partial_\tau\phi)
  -\frac{1}{2\tau}(\partial_\eta\phi)(\partial_\eta\phi)
  -\frac{\tau}{2} ({\bs\partial}_\perp\phi)\cdot({\bs\partial}_\perp\phi)
  -\frac{\tau}{2} m^2\phi^2
\end{equation}
including the measure $\sqrt{|g_{\mu\nu}|}=\tau$.  From this
Lagrangian, we see that the canonical momentum is given by
$\pi=\tau\partial_\tau\phi$.  Then the Hamiltonian density is
\begin{equation}
 \tau\Ham = \frac{1}{2\tau}\pi\pi +\frac{\tau}{2} \biggl[
  \frac{1}{\tau^2}(\partial_\eta\phi)(\partial_\eta\phi)
  +({\bs\partial}_\perp\phi)\cdot({\bs\partial}_\perp\phi)
  + m^2\phi^2\biggr] \;.
\end{equation}
Now that we have the Hamiltonian density, we can write down the time
independent Schr\"{o}dinger equation as follows,
\begin{equation}
 \int\rmd\eta\,\rmd^2 \x_\perp\biggl\{-\frac{1}{2\tau}
  \frac{\delta^2}{\delta\phi(\eta,\x_\perp)^2}
  +\frac{\tau}{2}\phi(\eta,\x_\perp)\Bigl[-
  \frac{\partial_\eta^2}{\tau^2} -{\bs\partial}_\perp^2 +
  m^2\Bigr]\phi(\eta,\x_\perp)\biggr\}\Psi_0 = E\Psi_0\; .
\end{equation}
The solution of the above Schr\"{o}dinger equation is then
\begin{equation}
 \begin{split}
 \Psi_0[\phiin] = & N\exp\biggl[-\frac{1}{2}\int\rmd\eta\,
  \rmd^2 \x_\perp\,\phiin(\eta,\x_\perp)\,
\tau\sqrt{-(1/\tau^2)\partial_\eta^2
  - {\bs\partial}_\perp^2+m^2}\,\phiin(\eta,\x_\perp)\biggr] \\
 = & N\exp\biggl[-\int\frac{\rmd k_\eta\,\rmd^2 \k_\perp}{(2\pi)^3}
  \phiin(-\k)\frac{\sqrt{(k_\eta/\tau)^2+\k_\perp^2+m^2}}{2\tau}
  \phiin(\k)\biggr]\; ,
\label{eq:wf_sc_tau}
 \end{split}
\end{equation}
where we defined the Fourier transform,
\begin{equation}
 \phi(\k) \equiv \int\rmd\eta\,\rmd^2 \x_\perp\,\tau\,
  \phi(\eta,x_\perp)\,\rme^{-\rmi(k_\eta \eta+\k_\perp\cdot \x_\perp)}\; ,
\end{equation}
with $\k\equiv(k_\eta,\k_\perp)$.  It is quite important to note that
this wavefunction is $\tau$ independent because the seeming $\tau$
dependence is to be absorbed by the variable change $\eta\to\eta/\tau$
in the first line and $k_\eta\to\tau k_\eta$ in the second line.  Of
course, such rescaling changes the argument of the fields
$\phiin(\eta,\x_\perp)$ and $\phiin(\k)$ too.  We will perform the
integration over those fields at the end, however, and so the
rescaling does not matter since possible $\tau$ dependence in
$\phiin(\eta,\x_\perp)$ and $\phiin(\k)$ would be absorbed in this
integration.  This feature is essential for our construction of the
wavefunction, for we made use of the time independent Schr\"{o}dinger
equation with an ansatz $\Psi=\rme^{-\rmi E\tau}\Psi_0$ where $\Psi_0$
is $\tau$ independent.


\section{Non-Abelian Gauge Theory}

  We are now ready to proceed into the construction of the
wavefunction in non-Abelian gauge theories. Before addressing the
calculation of the wavefunction, we summarize our conventions, gauge
choice, and basic equations.  Then we will construct the wavefunction
in the empty vacuum in the backward light cone region before the
collision.  We will connect the wavefunction from the infinitesimally
backward light cone region to the infinitesimally forward light cone
region with the boundary condition derived from the singularity in the
Hamiltonian density.


\subsection{Coordinates}

  Let us first recall our conventions for the light cone coordinates
and for the $\tau,\eta$ coordinates\footnote{Other definitions exist
in the literature, that differ mostly in the way the light-cone
coordinates are normalized.  This changes the detailed expression of
the metric tensor in light-cone coordinates.}.  The light cone
coordinates are defined by
\begin{equation}
 x^+ \equiv \frac{1}{\sqrt{2}}(t+z)\quad,\quad
 x^- \equiv \frac{1}{\sqrt{2}}(t-z)\; .
\end{equation}
{}From there, we can readily obtain the metric tensor from the
relation $\rmd s^2 = g_{\mu\nu}\rmd x^\mu\rmd x^\nu$,
\begin{equation}
g_{+-}=g_{-+}=1\quad,\quad g_{xx}=g_{yy}=-1\; ,
\end{equation}
where the non-written components are all zero.  Accordingly the light
cone derivatives are to be understood as
$\partial^\pm\equiv\partial/\partial x^\mp$.  

The proper time $\tau$ and the rapidity variable $\eta$ have already
been defined in Eq.~(\ref{eq:tau}), and are related to the light-cone
coordinates by $x^\pm =(\tau/\sqrt{2})\rme^{\pm\eta}$.  The metric
tensor for the $\tau,\eta$ coordinate system is diagonal, with
\begin{equation}
g_{\tau\tau}=1\quad,\quad g_{\eta\eta}=-\tau^2\quad,\quad
g_{xx}=g_{yy}=-1\; .
\end{equation}
Derivative with respect to $\tau,\eta$ and $x^{\pm}$ can be related
thanks to the following relation,
\begin{equation}
 \rmd x^\pm = \frac{x^\pm}{\tau}\rmd\tau \pm x^\pm\rmd\eta \; .
\end{equation}

{}From the above identities, we can write the one-form gauge
field\footnote{The transverse components have not been written because
they are the same in all the coordinate systems we consider here.} as,
\begin{equation}
 \begin{split}
 A_\tau\rmd\tau + A_\eta\rmd\eta &= A^-\rmd x^+ + A^+\rmd x^- \\
  &= \frac{1}{\tau}(x^+ A^- + x^- A^+)\rmd\tau
  +(x^+ A^- - x^- A^+)\rmd\eta \;.
 \end{split}
\end{equation}
Therefore, the components of the gauge field in the
$\tau,\eta$ coordinates are
\begin{equation}
 A_\tau = A^\tau \equiv \frac{1}{\tau}(x^+ A^- + x^- A^+)\quad,\quad
 A_\eta = -\tau^2 A^\eta \equiv x^+ A^- - x^- A^+\;.
\label{eq:def_eta}
\end{equation}
In the same way, we can obtain the field strength as the
two-form\footnote{The symbol $\wedge$ denotes the anti-symmetric
exterior product of forms.}  $F=\rmd A-\rmi g A\wedge A$, that is,
\begin{eqnarray}
&&
F_{\tau\eta} \equiv 
\partial_\tau A_\eta-\partial_\eta A_\tau -\rmi g[A_\tau,A_\eta]\; ,
\nonumber\\
&&F_{\tau i} \equiv 
\partial_\tau A_i-\partial_i A_\tau -\rmi g[A_\tau,A_i]\; ,
\nonumber\\
&&F_{\eta i} \equiv 
\partial_\eta A_i-\partial_i A_\eta -\rmi g[A_\eta,A_i]\; ,
\nonumber\\
&&F_{ij} \equiv 
\partial_i A_j-\partial_j A_i -\rmi g[A_i,A_j]\; ,
\end{eqnarray}
where $i$ refers to the perpendicular components ($x$ and $y$).  We
can raise or lower the indices using the metric tensor, e.g.
$F^{\tau\eta}=g^{\tau\tau}g^{\eta\eta}F_{\tau\eta}$.  One should keep
in mind, however, the fact that $F^{\tau\eta}$ is \textit{not} the
same as $\partial^\tau A^\eta-\partial^\eta A^\tau-\rmi
g[A^\tau,A^\eta]$.  In order to avoid confusion, we shall work
thoroughly in terms of quantities with lower indices, $\partial_\tau$,
$\partial_\eta$, $A_\tau$, and $A_\eta$.  The derivatives with the
present conventions are
\begin{equation}
 \partial_\tau=\partial^\tau \equiv \frac{\partial}{\partial\tau}\,,
  \quad
 \partial_\eta=-\tau^2\partial^\eta \equiv
  \frac{\partial}{\partial\eta}\,, \quad
 \partial_i=-\partial^i=\frac{\partial}{\partial x^i}\;.
\end{equation}


\subsection{Lagrangian and Hamiltonian}

  In the $\tau,\eta$ coordinates, the Lagrangian density is given by
\begin{equation}
 \tau\Lag =-\frac{\tau}{2}\tr\bigl[g^{\mu\nu}g^{\alpha\beta}
  F_{\mu\alpha}F_{\nu\beta}\bigr]  + \tau\Lag_{\text{source}}\;,
\end{equation}
where we have included the Jacobian\footnote{One has $\rmd^4x=\tau
\rmd\tau\rmd\eta\rmd^2\x_\perp$.} $\sqrt{|g_{\mu\nu}|}=\tau$ in the
definition of the density.  $\Lag_{\text{source}}$ represents the
singular source terms generated by the fast moving nuclei that collide
at $\tau=0$,
\begin{equation}
 \begin{split}
 & \Lag_{\text{source}} \equiv
   -\rho_1(\x_\perp)\delta(x^-)A^- -\rho_2(\x_\perp)\delta(x^+)A^+ \\
 =& -\frac{\tau}{2}\biggl[\frac{1}{x^+}\rho_1\,\delta(x^-)
  +\frac{1}{x^-}\rho_2\,\delta(x^+)\biggr]A_\tau
  -\frac{1}{2}\biggl[\frac{1}{x^+}\rho_1\,\delta(x^-)
  -\frac{1}{x^-}\rho_2\,\delta(x^+)\biggr]A_\eta \;.
 \end{split}
\label{eq:sources}
\end{equation}
The source $\rho_1(\x_\perp)$ represents the transverse color
distribution of the nucleus moving to the positive $z$ direction and
the source $\rho_2(\x_\perp)$ is the same quantity for the nucleus
moving in the negative $z$ direction. These densities are random
variables, and one should average observable quantities over the
statistical ensemble of these distributions. All the considerations we
make in the rest of this paper are for a given pair $\rho_1,\rho_2$
taken in this ensemble.

In the following discussion, we adopt $A_\tau=0$ as our gauge fixing
condition.  This gauge has an important benefit in the problem we want
to study. Indeed, one can see that it reduces to $A^+=0$ if $x+=0$ and
to $A^-=0$ if $x^-=0$. This implies that the color current associated
to the sources given in Eq.~(\ref{eq:sources}) is trivially conserved,
because the color field produced by the collision is such that the
terms that may have induced a precession of the current ($[A^+,J^-]$
or $[A^-,J^+]$) are identically zero. In practice, this means that we
can safely consider the color current as given once for all in terms
of $\rho_1$ and $\rho_2$, and forget about the current conservation
relation.

  In fact, naively, because $A^-\to0$ when $x^-\to0$ and $A^+\to0$
when $x^+\to0$, one may have the impression that the whole source term
in the Lagrangian density has no effect at all. However, this
conclusion is incorrect, because one has to differentiate with respect
to the fields in order to obtain the equations of motion, the latter
have a contribution from the source terms. Moreover, in the
$\tau,\eta$ coordinates, one has
\begin{equation}
 A_\eta\sim\mathcal{O}(\tau^2)\to 0
\label{eq:A_eta_zero}
\end{equation}
when $\tau\to0$, as one can see from the definition
(\ref{eq:def_eta}). In terms of the field component in these
coordinates, the Lagrangian density in the $A_\tau=0$ gauge reads
\begin{equation}
 \tau\Lag = -\frac{\tau}{2}\tr\biggl[-\frac{2}{\tau^2}(\partial_\tau
  A_\eta)^2 -2(\partial_\tau A_i)^2 +\frac{2}{\tau^2} F_{\eta i}^2
  +F_{ij}F_{ij}\biggr] + \tau\Lag_{\text{source}}\; ,
\end{equation}
and the canonical momenta are given by 
\begin{equation}
 \begin{split}
 E^i &\equiv \frac{\partial(\tau\Lag)}{\partial (\partial_\tau A_i)}
  =\tau\partial_\tau A_i = -\tau F_{i\tau}\;,\\
 E^\eta &\equiv \frac{\partial(\tau\Lag)}{\partial (\partial_\tau
  A_\eta)} =\frac{1}{\tau} \partial_\tau A_\eta
  = -\frac{1}{\tau} F_{\eta\tau}\;.
 \end{split}
\label{eq:momentum}
\end{equation}
As a consequence, the Hamiltonian density is
\begin{align}
 \tau\Ham =& 2\tr\bigl[(\partial_\tau A_i) E^i
  + (\partial_\tau A_\eta) E^\eta\bigr]-\tau\Lag \notag\\
 =& \tr\biggl[\frac{1}{\tau}E^i E^i + \tau E^\eta E^\eta
  +\frac{1}{\tau}F_{\eta i}F_{\eta i} +\frac{\tau}{2}F_{ij}F_{ij}
  \biggr] \notag\\
 &\qquad +\frac{\tau}{2}\biggl[\frac{1}{x^+}\rho_1\,\delta(x^-)
  -\frac{1}{x^-}\rho_2\,\delta(x^+)\biggr]A_\eta\;.
\end{align}
In the case of gauge theories, the time evolution is not uniquely
generated by the Hamiltonian due to the redundancy of the gauge
degrees of freedom.  It should become unique when the Gauss law
constraint, which is deduced from the coefficient of $A_\tau$ in the
Hamiltonian density (before the gauge condition is enforced),
\begin{equation}
 D_i E^i+D_\eta E^\eta-\frac{\tau^2}{2}\biggl[\frac{1}{x^+}\rho_1\,
  \delta(x^-)+\frac{1}{x^-}\rho_2\,\delta(x^+)\biggr] = 0\;,
\label{eq:Gauss}
\end{equation}
is satisfied by the physical state.  The last term of the Gauss law
looks like it may be zero (thanks to $\tau^2=2x^+x^-$), but we should
keep it because it contributes to the singularities of the Hamiltonian
when divided it by $\tau$, from which we can determine the boundary
conditions for the classical fields as shown in the next subsection.


\subsection{Classical background fields}

{}From the Hamiltonian, $H\equiv\int\rmd\eta\,\rmd^2 x_\perp\,\Ham$,
one gets the classical equations of motion for the classical
background fields $\Ac$ and momenta $\Ec$, which should also satisfy
the Gauss law (\ref{eq:Gauss}), as~\cite{roma}
\begin{align}
 \partial_\tau \Ec^i &= -\frac{\delta(\tau H)}{\delta \Ac_i}
  = \frac{1}{\tau}\D_\eta \Fc_{\eta i} + \tau \D_j \Fc_{ji}\; ,
\label{eq:eom_i} \\
 \partial_\tau \Ec^\eta &= -\frac{\delta(\tau H)}{\delta \Ac_\eta}
  = -\frac{1}{\tau}\D_j \Fc_{\eta j} -\frac{\tau}{2}
  \biggl[\frac{1}{x^+}\rho_1\,\delta(x^-)-\frac{1}{x^-}\rho_2\,
  \delta(x^+)\biggr]\; ,
\label{eq:eom_eta}
\end{align}
with the background covariant derivative $\D\equiv\partial-\rmi g\Ac$,
and the electric fields $\Ec^i=\tau\partial_\tau\Ac_i$ and
$\Ec^\eta=\tau^{-1}\partial_\tau\Ac_\eta$.  These equations must be
solved with retarded boundary conditions, in which the fields and
momenta are all vanishing when $t\to -\infty$. The solution of these
equations of motion is known analytically in the space-like regions
below the forward light-cone. Inside the forward light-cone, the
solution must be obtained numerically, with an initial condition at
$\tau=0^+$ which can be obtained analytically.  In order to determine
this initial condition, one must integrate the equations of motion
in an infinitesimal region just above the forward light-cone.

In terms of the sources $\rho_1$ and $\rho_2$, these initial
conditions are known to be~:
\begin{align}
 \Ac_i &= \alpha^1_i + \alpha^2_i \;,
\label{eq:cl_ai} \\
 \Ac_\eta &= 0 \;,
\label{eq:cl_aeta} \\
 \Ec^i &= \tau\partial_\tau\Ac_i= 0 \;,
\label{eq:cl_ei} \\
 \Ec^\eta &= \frac{1}{\tau}\partial_\tau \Ac_\eta
  = \rmi g[\alpha^1_i,\alpha^2_i] \;,
\label{eq:cl_eeta}
\end{align}
where $\alpha^1_i$ and $\alpha^2_i$ are known functions that depend
only on the perpendicular coordinates $\x_\perp$, and satisfy
$\partial_i\alpha^1_i=\rho_1$ and $\partial_i\alpha^2_i=\rho_2$.  (The
repeated index $i$ indicates a sum over the transverse indices $x$ and
$y$.)  In the rest of this subsection, we rederive these initial
conditions in the $\tau,\eta$ coordinates. This is instructive as it
illustrates some of the issues encountered when using the $\tau,\eta$
coordinates, and it will pave the way for the forthcoming construction
of the wavefunction.

\begin{figure}
 \begin{center}
 \includegraphics[width=6cm]{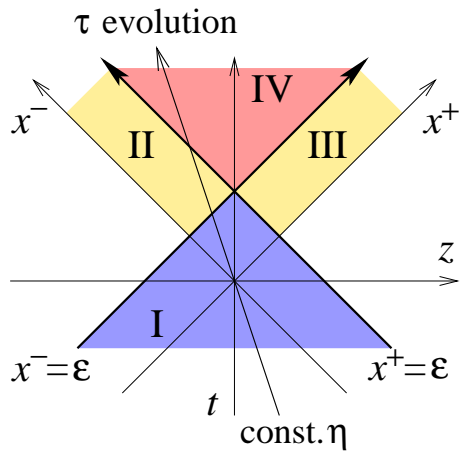}
 \caption{The source singularities are shifted by an infinitesimal
 amount $\epsilon$ in the $x^+$ and $x^-$ directions respectively.
 The $\tau$ integration from $0^-$ to $+\zeta$ is taken along a line
 of constant $\eta$.  The backward light cone region corresponds to a
 negative $\tau$.}
 \label{fig:evolve}
 \end{center}
\end{figure}

The main difficulty when dealing with the $\tau,\eta$ coordinates is
that the (physical) singularity of the sources is located at the same
position as the (unphysical) coordinate singularity\footnote{The
singularity of this system of coordinates at $\tau=0$ can be seen in
the fact that the determinant of the metric vanishes at $\tau=0$.} at
$\tau=0$. It should be noted that, while the backward light cone
region is parameterized by a negative $\tau$, the difficulty stems
from the impossibility to describe the space-like regions by a real
$\tau$.  We will introduce a regulator $\epsilon$, that we use to
slightly shift the location of the sources. This helps to separate the
physical singularities from the unphysical ones, as illustrated in
Fig.~\ref{fig:evolve}. Then the space-time is divided into four
regions,
\begin{displaymath}
 \begin{array}{lp{0mm}@:p{2mm}l}
  \text{Region I} &&& x^+<\epsilon,\quad x^-<\epsilon, \\
  \text{Region II} &&& 0<x^+<\epsilon,\quad x^->\epsilon, \\
  \text{Region III} &&& 0<x^-<\epsilon,\quad x^+>\epsilon, \\
  \text{Region IV} &&& x^+>\epsilon,\quad x^->\epsilon,
 \end{array}
\end{displaymath}
separated by the source singularities.  Let us consider the $\tau$
evolution along a line of constant $\eta$, e.g. in the negative $\eta$
region as illustrated in Fig.~\ref{fig:evolve}.  Then the $\tau$
evolution first goes through a singularity at $x^-=\epsilon$ from the
region I to the region II, and then through a second singularity at
$x^+=\epsilon$ that leads to the region IV.

  Equation (\ref{eq:cl_aeta}) is trivial from
Eq.~(\ref{eq:A_eta_zero}).  By inserting the canonical momenta into
the Gauss law divided by $\tau$, we can get the following boundary
condition,
\begin{equation}
 \int_{0^-}^\zeta\!\rmd\tau \biggl\{ \D_i\partial_\tau\Ac_i
  +\frac{1}{\tau^2}\partial_\eta\partial_\tau\Ac_\eta - \frac{\tau}{2}
  \biggl[\frac{1}{x^+}\rho_1\delta(x^--\epsilon) +\frac{1}{x^-}
  \rho_2\delta(x^+-\epsilon)\biggr]\biggr\} = 0 \;,
\label{eq:sing_Gauss}
\end{equation}
where we perform this integration (\ref{eq:sing_Gauss}) with respect
to $\tau$ in the small range from $0^-$ to $\zeta$ at fixed
$\eta$. The upper bound $\tau=\zeta$ should be chosen such that
$\zeta>\sqrt{2}\,\epsilon\,\rme^\eta$ and
$\zeta<\sqrt{2}\,\epsilon\,\rme^{-\eta}$, so that the integration ends
in the region II\footnote{Thanks to our shift of the sources by the
amount $\epsilon$, this path is entirely contained in the time-like
region, and can be parameterized by the $\tau,\eta$
coordinates.}. Since only the singularity at $x^-=\epsilon$ is present
on the integration path, we can assume that the background fields in
the region II are proportional to the step function
$\theta(x^--\epsilon)$. Note that such a step function gives a delta
function when differentiated with respect to $\tau$ or $\eta$.  Let us
therefore substitute the following forms,
\begin{equation}
 \Ac_i = \alpha^1_i\theta(x^--\epsilon)\quad,\quad
 \Ac_\eta = \tau^2\beta^1\theta(x^--\epsilon)\;,
\end{equation}
into Eq.~(\ref{eq:sing_Gauss}). The only terms that survive in this
integral in the limit $\epsilon\to 0^+$ are those containing a delta
function.  It should be noted that the $\Ac_i$ in the covariant
derivative $\D_i$ drops out because it has the same color structure as
$\partial_\tau\Ac_i$, as long as only the singular part is concerned
(obtained when $\partial_\tau$ acts on the step function rather than
on $\alpha^1_i$).  After all, we get
\begin{equation}
 \partial_i\alpha^1_i - 2\beta^1 - \rho_1 = 0\; .
\label{eq:cond_I}
\end{equation}
The equation of motion (\ref{eq:eom_i}) is then satisfied almost
trivially, and the other equation of motion (\ref{eq:eom_eta}) leads
to the very same boundary condition, and therefore we cannot determine
uniquely $\alpha_i^1$ and $\beta_1$.  This degeneracy is related to
the residual gauge freedom; if we allow a nonzero $\beta^1$
proportional to $\rho_1$, then we can eliminate it by a gauge
rotation\footnote{This situation is actually analogous to what happens
when solving the equations of motion in the light-cone coordinates in
the $A^+=0$ gauge, in the presence of the source singularity
$\delta^{\mu+}\rho_1\delta(x^-)$. In this case, $A^-$ can be gauged
away by an $x^+$ independent gauge rotation.}  applied to $\Ac_\eta$
without affecting the gauge condition $\Ac_\tau=0$ (except for terms
of order $\tau\to0$).  We shall choose $\beta^1=0$ and therefore
$\partial_i\alpha^1_i=\rho_1$. We can do the same for the region III,
where we obtain the background $\Ac_i=\alpha^2_i\theta(x^+-\epsilon)$
with $\partial_i\alpha^2_i=\rho_2$.

Then, we shall push the time integration towards the region IV.\ \
There is another singularity at $x^+=\epsilon$ for $\eta<0$ and at
$x^-=\epsilon$ for $\eta>0$, which can be accounted for by using the
following parameterization,
\begin{equation}
 \Ac_i=\alpha^1_i\theta(x^--\epsilon)\theta(\epsilon-x^+)
  +\alpha^2_i\theta(x^+-\epsilon)\theta(\epsilon-x^-)
  +\alpha^3_i\theta(x^--\epsilon)\theta(x^+-\epsilon)\;,
\end{equation}
where $\alpha^1_i$ and $\alpha^2_i$ are already fixed. If we
differentiate twice with respect to $\tau$ or $\eta$ the products of
theta functions in this expression, we obtain terms with two delta
functions that are singular at $\tau=\eta=0$ (in the limit
$\epsilon\to 0^+$).  From the equation of motion (\ref{eq:eom_i}), we
therefore get the condition,
\begin{equation}
 \int_{0^-}^{0^+}\rmd\eta\,\int_{0^-}^\zeta\rmd\tau\,
  \frac{1}{\tau}\partial_\eta^2\Ac_i = 0 \;,
\label{eq:eta2zero}
\end{equation}
where this time we must chose
$\zeta>\sqrt{2}\,\epsilon\,\rme^{\pm\eta}$ so that the endpoint of the
$\tau$ integration is in the region IV.\ \ This relation leads to
\begin{equation}
 \alpha^3_i = \alpha^1_i + \alpha^2_i \;.
\label{eq:alpha3}
\end{equation}
Let us finally perform the integration (\ref{eq:sing_Gauss}) with the
ansatz
\begin{equation}
 \Ac_\eta = \tau^2\beta^3\theta(x^+-\epsilon)\theta(x^--\epsilon)\; .
\end{equation}
In this case, it is important to note that $\partial_\tau\Ac_i$ can
have a different color structure than that of $\Ac_i$. Therefore, we
should keep the covariant derivative $\D_i$ and treat carefully the
commutators such as $[\Ac_i,\partial_\tau\Ac_i]$ and
$[\Ac_i,\partial_\eta\Ac_i]$.  The Gauss law (\ref{eq:sing_Gauss})
eventually leads to the boundary condition
\begin{equation}
 \partial_i\alpha^1_i + \partial_i\alpha^2_i
  -\rmi g[\alpha^1_i,\alpha^2_i] +2\beta^3 - \rho_1 - \rho_2 = 0\; ,
\end{equation}
and the equation of motion (\ref{eq:eom_eta}) gives again the same
condition, which results in
\begin{equation}
 \beta^3=\frac{\rmi g}{2}\; [\alpha^1_i,\alpha^2_i]\; .
\end{equation}

Hence, one can reproduce the classical background fields
(\ref{eq:cl_ai})-(\ref{eq:cl_eeta}) in the $\tau,\eta$ coordinates by
slightly shifting the source singularities.  We should emphasize here
that, even though we have $\Ec^i=0$, we must keep terms such as
$\partial_\tau\Ec^i$ and $\tau^{-1}\Ec^i$ which are non-vanishing when
$\tau\to 0$. Since $\Ac_\eta$, in contrast, goes to zero like
$\mathcal{O}(\tau^2)$, we can drop $\partial_\tau \Ac_\eta=0$ and
$\tau^{-1}\Ac_\eta=0$ in the limit $\tau\to0$.


\subsection{Wavefunction in the backward light cone region}

  We now proceed with the derivation of the wavefunction in the
$\tau,\eta$ coordinates in the non-Abelian gauge theory.  Let us first
consider the backward light cone region in the weak coupling regime.
We see that $A_\eta$ and $E^\eta$ are not a dynamical variables on the
light cone in the $A_\tau=0$ gauge.  Therefore, the wavefunction does
not involve $A_\eta$ at all and the Gauss law provides $E^\eta$ as a
function of the dynamical variables.  In other words, $E^\eta$ is a
solution of the Gauss law with $A_\eta=0$ substituted,
\begin{equation}
 E^\eta = -\frac{1}{\partial_\eta}D_i E^i\; ,
\label{eq:Gauss_eeta}
\end{equation}
where we should remark that the fields and the momenta in this
identity are the full ones, including both the classical backgrounds
and the quantum fluctuations.

In the following, we denote the quantum fluctuations with lowercase
letters,
\begin{equation}
 a_i \equiv A_i-\Ac_i\;,\quad a_\eta \equiv A_\eta-\Ac_\eta\;, \quad
 e^i \equiv E^i-\Ec^i\;,\quad e^\eta \equiv E^\eta-\Ec^\eta\; .
\label{eq:fluctuation}
\end{equation}
We then expand the Hamiltonian to quadratic order in the fluctuations,
and then the time-independent Schr\"{o}dinger equation can be written
as
\begin{equation}
 \begin{split}
  & \int\rmd\eta\,\rmd^2\x_\perp\,\tr\biggl\{-\frac{1}{\tau}
  \frac{\delta^2}{\delta a_i^2}-\tau\frac{\delta^2}{\delta a_\eta^2}
  +\frac{1}{\tau}(\partial_\eta a_i-\D_i a_\eta)^2 \\
  & \qquad +\frac{\tau}{2}\Bigl[(\D_i a_j-\D_j a_i)^2 -2\rmi g
  \Fc_{ij}[a_i,a_j]\Bigr] \biggr\}\Psi_-[A]  = E\Psi_-[A]\; ,
\label{eq:sch_back}
 \end{split}
\end{equation}
with an energy eigenvalue $E$. When writing this Schr\"{o}dinger
equation, we have replaced the canonical momenta by differential
operators, thanks to the canonical commutation relations at equal
$\tau$,
\begin{align}
 \bigl[a_i^a(\eta,\x_\perp),e^{jb}(\eta',\x_\perp')\bigr]
  &=\rmi \delta^{ab}
  \delta^{ij}\delta(\eta-\eta')\delta^{(2)}(\x_\perp-\x_\perp')\;, \\
 \bigl[a_\eta^a(\eta,\x_\perp),e^{\eta b}(\eta',\x_\perp')\bigr] &=\rmi
  \delta^{ab}\delta(\eta-\eta')\delta^{(2)}(\x_\perp-\x_\perp')\;.
\end{align}
It should be mentioned here that the right hand side does not involve
$\tau$ explicitly because we have defined the canonical momenta $e^i$
and $e^\eta$ from $\tau H$, including the measure.

  Changing temporarily to the variables $a_\zeta\equiv a_\eta/\tau $
(and $\partial_\eta\equiv\tau\partial_\zeta$), and using the notation
$a_{_I}\equiv(a_i,a_\zeta)$, we can write the Schr\"{o}dinger equation
in the following compact form,
\begin{equation}
 \int\rmd\eta\,\rmd^2\x_\perp\,\tr\biggl[-\frac{1}{\tau}
  \frac{\delta^2}{\delta a_{_I}^2} + \tau\,a_{_I}\cdot G^{-1}_{_{IJ}}
  \cdot a_{_J}\biggr]\Psi_-[A] = E\Psi_-[A]\;,
\label{eq:sch_back2}
\end{equation}
where we have defined the background propagator,
\begin{equation}
 (G^{-1}_{_{IJ}})^{ab} \equiv -\delta_{_{IJ}}(\D^2)^{ab}
 + (\D_{_I}\D_{_J})^{ab} -2gf^{acb}\Fc_{_{IJ}}^c\; ,
\label{eq:back_propagator}
\end{equation}
with $\D^2\equiv \partial_\zeta^2+\D_i\D_i$.  The formal solution to
this Schr\"{o}dinger equation is not difficult to find.  The ground
state solution to Eq.~(\ref{eq:sch_back}) or Eq.~(\ref{eq:sch_back2})
can be expressed in terms of the square-root of the inverse propagator
$G^{-1}$ as
\begin{equation}
 \Psi_- = N\exp\biggl[-\int\rmd\eta\,\rmd^2 \x_\perp\,\tau\,\tr\Bigl(
  a_{_I}\,\sqrt{G^{-1}_{_{IJ}}}\,a_{_J} \Bigr)\biggr]\; ,
\end{equation}
where we have chosen only the normalizable solution.  Since there is
no background field in the backward light cone region, the square-root
of the inverse propagator is easy to calculate. It is simply
proportional to the transverse projection operator
$\delta_{_{IJ}}-\partial_{_I}\partial_{_J}/\partial^2$.  Noting that
$a_\zeta=\tau^{-1}a_\eta\to0$, we see that the wavefunction is simply
\begin{equation}
 \Psi_-[A]=N\exp\Biggl\{\!-\!\!\int\!\rmd\eta\,\rmd^2 \x_\perp\,\tr
  \biggl[a_i\,\tau\sqrt{\!-\left(\frac{\partial_\eta}{\tau}\right)^2
  \!-\!{\bs\partial}_\perp^2}\,\Bigl(\delta_{ij}
  -\frac{\partial_i\partial_j}{(\frac{\partial_\eta}{\tau})^2
  +{\bs\partial}_\perp^2}\Bigr) a_j\biggr]\Biggr\}\;.
\label{eq:wf_back}
\end{equation}
Except when $\partial_\eta$ is smaller than $\tau Q_s$, the
$\partial_\eta/\tau$ term dominates over the transverse derivative.
For the analysis of the instability, however, we may have to retain
the transverse momentum here, because the unstable $\eta$ dependent
modes may be very soft~\cite{roma}.  Also we note that, although we
have used the notation $a_i$, it is identical to the full fields $A_i$
in the absence of backgrounds in the backward light cone region.


\subsection{Singularity in the Hamiltonian}

  In order to clarify the boundary condition at the source
singularity, we first need the Hamiltonian in the presence of
background fields, and more specifically its singularity.  Note
that the singular part of the Hamiltonian should commute with the
Gauss law because of gauge invariance.

The explicit form of the Hamiltonian in the presence of the
background reads
\begin{align}
 & \tau\Ham = \tr\biggl[\frac{1}{\tau}e^i e^i +\frac{2}{\tau}\Ec^i e^i
  +\frac{1}{\tau}\Ec^i\Ec^i +\tau e^\eta e^\eta + 2\tau\Ec^\eta e^\eta
  + \tau \Ec^\eta\Ec^\eta \notag\\
 & +\frac{1}{\tau}\bigl(\partial_\eta a_i -\D_i a_\eta)^2
  -\frac{2}{\tau}\rmi g\,(\partial_\eta\Ac_i) [a_\eta,a_i]
  +\frac{2}{\tau}(\partial_\eta\Ac_i) (\partial_\eta a_i -\D_i a_\eta)
  +\frac{1}{\tau}(\partial_\eta\Ac_i)^2 \notag\\
 & \quad+\frac{\tau}{2}\bigl(\D_i a_j-\D_j a_i\bigr)^2 -\tau\rmi g\,
 \Fc_{ij}[a_i,a_j] + \tau\,\Fc_{ij}(\partial_i a_j-\partial_j a_i)
   +\frac{\tau}{2}\Fc_{ij}^2 \notag\\
 & \qquad+\tau\Bigl\{\frac{1}{x^+}\rho_1\,\delta(x^-)
  -\frac{1}{x^-}\rho_2\,\delta(x^+)\Bigr\}a_\eta\biggr]\; ,
\end{align}
up to quadratic order in the fluctuations.  The last term is the
current density due to the sources, which can also be expressed as
$J^\eta=-(2/\tau)\partial_i\partial_\eta\Ac_i$.  Gauge invariance
requires the covariant conservation of the current, i.e. $D_\eta
J^\eta=0$, which is actually satisfied thanks to
Eq.~(\ref{eq:eta2zero}), since $A_\eta\to0$ at $\tau\to0$ .

  Thus, in the limit $a_\eta\to0$, the singular terms in the
Hamiltonian are
\begin{equation}
 \tau\Ham_{\text{sing}} = \tr\biggl[\,2(\partial_\tau\Ac_i) e^i
  + \frac{2}{\tau}(\partial_\eta\Ac_i)
  (\partial_\eta a_i) +\frac{1}{\tau}(\partial_\eta\Ac_i)^2\biggr]\; .
\end{equation}
Here we have used Eq.~(\ref{eq:momentum}) in order to rearrange the
first term.  The delta-function singularities originate from
$\partial_\eta$ and $\partial_\tau$ acting on the step function in
$\Ac_i$.  The second and third terms can be in fact dropped thanks to
Eq.~(\ref{eq:eta2zero}), after an integration by parts.  Therefore,
the only remaining singular term is after all,
\begin{equation}
 \tau\Ham_{\text{sing}} = 2\tr\bigl[\,(\partial_\tau \Ac_i)
  e^i\bigr]\; ,
\label{eq:ham_sing}
\end{equation}
on the light cone where $a_\eta$ is vanishing.


\subsection{In the forward light cone region}

  By integrating the Schr\"{o}dinger equation with respect to $\tau$
from $0^-$ to $\zeta$ with the singularity in the Hamiltonian we have
just identified in Eq.~(\ref{eq:ham_sing}), we can derive a relation
between the wavefunctions in the infinitesimally backward and forward
light cone regions.  

  In principle, we can follow the same procedure as the one used for
the determination of the classical backgrounds.  By shifting the
singularities as shown in Fig.~\ref{fig:evolve}, we can access the
space-like regions while using the $\tau,\eta$ coordinates.  In the
presence of the regulator $\epsilon$, we should keep $A_\eta$ terms
which are of order $\epsilon^2$, and the $\tau$ integration from I
towards IV will pick up singular terms proportional to $A_\eta$.  When
the wavefunction contains linear terms in $A_\eta$ in the exponential,
the $A_\eta$ integration in the Wigner transformation results in the
delta-function constraint imposed on the conjugate variable $E^\eta$.
At the end in the limit of $\epsilon\to0$, we must take simultaneously
the limit $A_\eta\to0$, so that the $A_\eta$ terms in the wavefunction
disappear.  Nevertheless the constraint on $E^\eta$ preserves some
information about the singular terms.  This procedure is a bit tedious
to implement, but the final answer is quite simple: the Gauss law
always holds, from which one can obtain $E^\eta$ as a function of the
other fields.  Therefore, the constraint on $E^\eta$ after the
$A_\eta$ integration simply coincides with the Gauss law
(\ref{eq:Gauss}).  This means that we can obtain the final answer by
the following shortcut: drop all the singular terms involving
$A_\eta$ and add by hand to the Wigner function a delta-function
constraint fixing $E^\eta$ from the Gauss law.

  Since we can set $A_\eta=0$ even with finite $\epsilon$, the
singular Hamiltonian (\ref{eq:ham_sing}) is sufficient for our purpose
to connect the wavefunctions between the two light-cones.  The
Schr\"{o}dinger equation leads to the boundary condition from
$\tau=0^-$ to $\tau=\zeta\to 0^+$,
\begin{equation}
 \rmi (\Psi_+ - \Psi_-) = \Bigl(\int_{0^-}^\zeta\!\rmd\tau
  \int\rmd\eta\,\rmd^2 \x_\perp\,\tau
  \Ham_{\text{sing}}\Bigr)\frac{1}{2}(\Psi_+ + \Psi_-)\; .
\end{equation}
Neglecting higher order terms in the fluctuations, we can solve the
above equation as
\begin{equation}
 \Psi_+ \simeq \exp\biggl[-\rmi\int_{0^-}^\zeta\!\rmd\tau
  \int\rmd\eta\,\rmd^2 \x_\perp\,
  \tau\Ham_{\text{sing}}\biggr]\Psi_-\; ,
\end{equation}
where $\tau\Ham_{\text{sing}}$ is given by Eq.~(\ref{eq:ham_sing}).
The identification of $e^i$ as the differential operator
$-\rmi\delta/\delta A_i$ enables us to write
\begin{equation}
 \int_{0^-}^\zeta\!\rmd\tau\,\tau\Ham_{\text{sing}}
  =2\tr\bigl[\Ac_i(\zeta)\, e^i\bigr]
  =-\rmi \Ac_i^a(\zeta) \frac{\delta}{\delta A_i^a}\; .
\end{equation}
The exponential of this differential operator is nothing but the
translation operator of $A_i$ by an amount $-\Ac_i$~; i.e. $A_i\to
A_i-\Ac_i$.  Therefore, we conclude that $\Psi_+[A]=\Psi_-[A-\Ac]$.
By taking the limit $\epsilon\to 0^+$ while keeping the final time
$\zeta$ very small but fixed, the surface of constant time
$\tau=\zeta$ lies entirely in the region IV, where the classical
transverse field is $\Ac_i=\alpha^1_i+\alpha^2_i$. Thus, we finally
obtain the following wavefunction in the forward light-cone,
\begin{equation}
 \Psi_+[A]=\Psi_-[A-(\alpha^1+\alpha^2)]\; ,
\end{equation}
with the wavefunction defined in Eq.~(\ref{eq:wf_back}).

  Although the wavefunction takes the same functional form, it is the
Gauss law which is actually affected nontrivially by the source
singularity, and the fluctuations in $e^\eta$ are uniquely fixed from
the background fields (\ref{eq:cl_ai}) and (\ref{eq:cl_eeta}), and the
fluctuations $a_i$ and $e^i$.  Namely, from Eq.~(\ref{eq:Gauss_eeta}),
up to term of quadratic order in fluctuations, we have
\begin{equation}
 e^\eta = e^\eta_0 -\int_{\eta_0}^\eta\rmd\eta\;\D_i e^i \; ,
\end{equation}
where we assumed that the fluctuation $e^i$ does not contain a
singularity of the type $\sim\tau\partial_\tau\theta(x^\pm-\epsilon)$
(unlike the classical transverse electric field $\Ec^i$ -- The $\eta$
integration of $\D_i\Ec^i$ leads to $\Ec^\eta$ because of this
singularity.)  The origin of the presence of an indefinite integral
constant $e^\eta_0$ is that the Gauss law cannot constrain the $\eta$
zero-mode of $e^\eta$.  In such a case, the $a_\eta$ integration in
the Wigner transform would impose the zero-mode to vanish.  Hence,
$e^\eta_0$ should be fixed by the condition
$\int_{-\infty}^\infty\!\rmd\eta\,e^\eta=0$.


\subsection{Wigner transformation}
Finally, we can compute the Wigner function from the wavefunction.
The conditions of $a_\eta\to0$ and the Gauss law constraint are, as we
have noted before, expressed by delta-functions added by hand.  We can
then write the Wigner function as follows~:
\begin{align}
 & W[a,e] \notag\\
 &= N\int\big[\rmd\delta a\big]\;\Psi_+^\ast[a-\half\delta a]
 \, \Psi_+[a+\half\delta a]\,\rme^{-\rmi\int\rmd\eta\,\rmd^2 \x_\perp\,
  2\tr[e^i \delta a_i]} \notag\\
 &= N\exp\Biggl\{-\int\rmd\eta\,\rmd^2 \x_\perp\,2\tr\biggl[a_i\,\tau
  \sqrt{-(\partial_\eta/\tau)^2-{\bs\partial}_\perp^2}\,\Bigl(\delta_{ij}
  -\frac{\partial_i \partial_j}
  {(\partial_\eta/\tau)^2+{\bs\partial}_\perp^2}\Bigr)\,a_j \notag\\
 &\qquad +e^i\frac{1}{\tau\sqrt{-(\partial_\eta/\tau)^2
  -{\bs\partial}_\perp^2}}\,\Bigl(\delta_{ij}+\frac{\partial_i \partial_j}
  {(\partial_\eta/\tau)^2}\Bigr)\, e^j\biggr]\Biggr\} \notag\\
 &\qquad\qquad\times\delta[a_\eta]\,\delta\biggl[e^\eta-e^\eta_0+\int_{\eta_0}^\eta\rmd\eta\;\D_i e^i\biggr]\;,
\label{eq:wigner}
\end{align}
where $\delta_{ij}+\partial_i\partial_j/(\partial_\eta/\tau)^2$ is the
inverse of $\delta_{ij}-\partial_i\partial_j/[(\partial_\eta/\tau)^2
+{\bs\partial}_\perp^2]$.  This Wigner function allows us to compute
the average of any observable $\mathcal{O}[A]$ at later time $\tau>0$
with initial quantum fluctuations, as follows\footnote{A result very
similar to ours was already known in the case of a scalar field
subject to a parametric resonance \cite{son2}; namely that calculating
the expectation value of some combination of fields can be done with a
solution of the classical equation of motion, provided one averages
with a Gaussian weight over fluctuations of the initial condition for
this classical solution.  See also Ref.~\cite{jeon}.},
\begin{equation}
 \langle\mathcal{O}\rangle_\tau = \int\big[\rmd a_i\,\rmd a_\eta\,\rmd
  e^i\,\rmd e^\eta\big]\;W[a,e]\;\mathcal{O}\bigl[
  \Ac[\Ac_i+a_i,a_\eta,e^i,\Ec^\eta+e^\eta;\tau]\bigr]\;.
\label{eq:formula_QCD}
\end{equation}
Here we symbolically denote by
$\Ac[\Ac_i+a_i,a_\eta,e^i,\Ec^\eta+e^\eta;\tau]$ the
classical gauge field  obtained at time $\tau$ by solving the
classical equations of motion (\ref{eq:eom_i}) and (\ref{eq:eom_eta})
with initial conditions $\Ac_i+a_i$, $a_\eta$, $e^i$,
and $\Ec^\eta+e^\eta$ at $\tau=0^+$, where $\Ac_i$ and $\Ec^\eta$ are
the usual background fields (\ref{eq:cl_ai}) and
(\ref{eq:cl_eeta}) respectively.

  The Gaussian Wigner function describes the dis\-per\-sion of
fluc\-tua\-tions around the classical initial condition.  The
distribution turns out not to be of white-noise type but to be
characterized by the correlation function,
\begin{align}
 & \bigl\langle a_i(\eta,\x_\perp)\,a_j(\eta',\x_\perp')
 \bigr\rangle
 = \frac{\displaystyle \int\big[\rmd a_i\,\rmd a_\eta\,\rmd e^i\,
  \rmd e^\eta\big]\;W[a,e]\;a_i(\eta,\x_\perp)\,a_j(\eta',\x_\perp')}
  {\displaystyle \int\big[\rmd a_i\,\rmd a_\eta\,\rmd e^i\,\rmd e^\eta\big]\;
  W[a,e]} \notag\\
 &= \frac{1}{\tau\sqrt{-(\partial_\eta/\tau)^2-{\bs\partial}_\perp^2}}
  \Bigl(\delta_{ij}+\frac{\partial_i\partial_j}
  {(\partial_\eta/\tau)^2}\Bigr)\,\delta(\eta-\eta')\,
  \delta^{(2)}(\x_\perp\!-\!\x_\perp') \;.
\label{eq:dis_a}
\end{align}
The fluctuation of canonical momenta is characterized inversely by
\begin{equation}
 \begin{split}
 & \bigl\langle e^i(\eta,\x_\perp)\,e^j(\eta',\x_\perp')
  \bigr\rangle \\
 &= \tau\sqrt{-(\partial_\eta/\tau)^2
  -{\bs\partial}_\perp^2}\Bigl(\delta_{ij}-\frac{\partial_i\partial_j}
  {(\partial_\eta/\tau)^2+{\bs\partial}_\perp^2}\Bigr)\,
  \delta(\eta-\eta')\,\delta^{(2)}(\x_\perp\!-\!\x_\perp') \; .
 \end{split}
\label{eq:dis_e}
\end{equation}
The formula (\ref{eq:formula_QCD}) and the above two relations,
Eqs.~(\ref{eq:dis_a}) and (\ref{eq:dis_e}), are our central result.
Regarding the $\eta$ components, $a_\eta$ is vanishing while $e^\eta$
is fixed by the Gauss law.  In the limit of $\tau\to0$ the
distribution becomes simpler, and depends solely on the derivative
with respect to rapidity, $\partial_\eta$.  For practical uses,
however, it is not possible to start the numerical simulation exactly
at $\tau=0^+$, and the transverse derivatives ${\bs\partial}_\perp$
will then play the role of a regulator for infrared modes when
$\partial_\eta/\tau$ is smaller than ${\bs\partial}_\perp$.


\section{Conclusions}

  In this paper, we have discussed an approximation scheme for
calculating the quantum expectation value of an observable that
depends on field configurations at late times.  In the WKB
approximation, we have obtained a formula giving this expectation
value from the classical equations of motion, with an initial
condition that includes fluctuations.  The spectrum of these initial
fluctuations is specified by the Wigner transform of the wavefunction,
the latter being obtained as a ground state solution of the
Schr\"{o}dinger equation.  Our formula is generic enough that we can
expect a variety of applications, for instance to the study of an
unstable system, or to that of the spinodal decomposition associated
with a phase transition.

An application of particular interest to heavy ion collisions is to
calculate the Wigner function of these fluctuations immediately after
the initial impact of the two nuclei, in the weak coupling regime.
These initial quantum fluctuations provide the seeds that trigger the
instability of the rapidity dependent modes.  The precise
determination of initial fluctuations or seeds is of crucial
importance for the estimate of the growth rate of the unstable modes.

By identifying the physically relevant gauge fluctuations and the
singularity in the Hamiltonian density, we have derived the
discontinuity between the wavefunctions in the backward and forward
light-cones.  It turns out that the initial singularity results only
in a shift in the transverse gauge fields, equal to the classical
background field.  This is because, on the light-cone, the physical
degrees of freedom are gauge fields polarized in the perpendicular
plane and the longitudinal mode along the collision axis has to
vanish.

This wavefunction and the resultant Wigner function describe the
spectrum of initial fluctuations. There are two noteworthy features
in our results.

Firstly, both $e^i$ and $a_i$ fluctuate, with 2-point correlations that
are inversely related to one another.  This means, if the boost
invariance is largely violated by $e^i$, then $a_i$ fluctuations are
suppressed, and vice versa.  Thus, the prescription adopted in
Ref.~\cite{roma}, in which only the $e^i$ fluctuations are taken into
account, needs to be extended in order to include also the
fluctuations of $a_i$.  It should be important to describe these
fluctuations correctly because the $a_i$ distribution rather than
$e^i$ distribution has large correlation for infrared modes having
a small rapidity dependence.

Secondly, although it may be less important, the off-diagonal
correlation between $x$ and $y$ components is not vanishing. Again,
this might be relevant for infrared modes.  This feature is also
missing in Ref.~\cite{roma}.

  It would be interesting to see how these modifications of the
fluctuation spectrum affect the growth rate of the instability.
Although the spectrum of the initial fluctuations could be obtained
analytically, this question can only be investigated numerically,
since it requires to solve the classical Yang-Mills equations in the
forward light-cone.


\section{Acknowledgments}

  We thank Raju Venugopalan and Tuomas Lappi for useful discussions
and Sangyong Jeon for comments.  This work was supported in part by
the RIKEN BNL Research Center and the U.S.\ Department of Energy under
cooperative research agreement \#DE-AC02-98CH10886.  FG would like to
thank the Institute of Nuclear Theory at the University of Washington
for its hospitality and financial support while this work was being
completed.


\end{document}